# Stratospheric Balloons as a Complement to the Next Generation of Astronomy Missions

P. Maier[a]*, M. Ångermann[d], J. Barnstedt[e], S. Bougueroua[a], A. Colin[c], L. Conti[c], R. Duffard[c], L. Hanke[e], O. Janson[d], C. Kalkuhl[e], N. Kappelmann[e], T. Keilig[a], S. Klinkner[a], A. Krabbe[a], M. Lengowski[a], C. Lockowandt[d], T. Müller[b], J.-L. Ortiz[c], A. Pahler[e], T. Rauch[e], T. Schanz[e], B. Stelzer[e], M. Taheran[a], A. Vaerneus[d], K. Werner[e], J. Wolf[a]

[a] *Institute of Space Systems, University of Stuttgart, Pfaffenwaldring 29, 70569 Stuttgart*, Germany, pmaier@irs.uni-stuttgart.de
[b] *Max-Planck-Institut für extraterrestrische Physik, Gießenbachstraße, 85741 Garching, Germany*
[c] *Instituto de Astrofísica de Andalucía, Glorieta de la Astronomía S/N, 18008 Granada, Spain*
[d] *Swedish Space Corporation, Torggatan 15, 17154 Solna, Sweden*
[e] *Institut für Astronomie und Astrophysik, Universität Tübingen, Sand 1, 72076 Tübingen, Germany*
* Corresponding Author

## Abstract

Observations that require large physical instrument dimensions and/or a considerable amount of cryogens, as it is for example the case for high spatial resolution far infrared astronomy, currently still face technological limits for their execution from space. The high cost and finality of space missions furthermore call for a very low risk approach and entail long development times.

For certain spectral regions, prominently including the mid- to far-infrared as well as parts of the ultraviolet, stratospheric balloons offer a flexible and affordable complement to space telescopes, with short development times and comparatively good observing conditions. Yet, the entry burden to use balloon-borne telescopes is high, with research groups typically having to shoulder part of the infrastructure development as well. Aiming to ease access to balloon-based observations, we present the efforts towards a community-accessible balloon-based observatory, the European Stratospheric Balloon Observatory (ESBO). ESBO aims at complementing space-based and airborne capabilities over the next 10-15 years and at adding to the current landscape of scientific ballooning activities by providing a service-centered infrastructure for broader astronomical use, performing regular flights and offering an operations concept that provides researchers with a similar proposal-based access to observation time as practiced on ground-based observatories.

We present details on the activities planned towards the goal of ESBO, the current status of the STUDIO (Stratospheric UV Demonstrator of an Imaging Observatory) prototype platform and mission, as well as selected technology developments with extensibility potential to space missions undertaken for STUDIO.

## 1. Acronyms/Abbreviations

| | |
|---|---|
| ACU | Attitude Control Unit |
| ALMA | Atacama Large Millimeter/submillimeter Array |
| ASTHROS | Astrophysics Stratospheric Telescope for High Spectral Resolution Observations at Submillimeter-wavelengths |
| BLAST | Balloon-borne Large Aperture Submillimeter Telescope |
| CAFÉ | Census of WHIM Accretion Feedback Explorer |
| CFRP | Carbon Fibre Reinforced Polymer |
| COTS | Commercial Off-The-Shelf |
| ECSS | European Cooperation for Space Standardization |
| ESA | European Space Agency |
| ESBO *DS* | European Stratospheric Balloon Observatory *Design Study* |
| FIR | Far Infrared |
| FPGA | Field Programmable Gate Array |
| IAA | Indian Institute of Astrophysics |
| IAAT | Institut für Astronomie und Astrophysik Tübingen |
| IRAS | Infrared Astronomical Satellite |
| ISS | Image Stabilization System |
| JWST | James Webb Space Telescope |
| MCP | Microchannel Plate |
| NIR | Near Infrared |
| NOEMA | Northern Extended Millimeter Array |
| ORFEUS | Orbiting and Retrievable Far and Extreme Ultraviolet Spectrometer |
| ORISON | innOvative Research Infrastructure based on Stratospheric balloONs |
| OS | Operating System |
| OSS | Origins Survey Spectrograph |
| OST | Origins Space Telescope |
| PILOT | Polarized Instrument for the Long- |

| | wavelength Observation of the Tenuous ISM |
|---|---|
| PUS | Packet Utilization Standard |
| SOFIA | Stratospheric Observatory For Infrared Astronomy |
| SPICA | Space Infrared Telescope for Cosmology and Astrophysics |
| SSC | Swedish Space Corporation |
| STO | Stratospheric Terahertz Observatory |
| STUDIO | Stratospheric UV Demonstrator of an Imaging Observatory |
| SuperBIT | Superpressure Balloon-borne Imaging Telescope |
| TINI | Tübingen IAA Nebular Investigator |
| TIP | Telescope Instruments Platform |
| UV | Ultraviolet |
| VIS | Visible |
| WD | White Dwarf |
| ZPB | Zero Pressure Balloon |

# 1. Introduction

## 1.1 The current situation of balloon-borne astronomy

The idea of using stratospheric balloons to overcome the obstructions of Earth's atmosphere for astronomical observations is not new. Historically, the advantages were obvious: spacecraft did not exist or were hardly available and capabilities of planes were limited, leaving balloons as the only option to move instruments above most of the atmosphere. In current times, the benefits do not seem as clear: both spacecraft and planes provide powerful observation platforms and ground-based telescopes invest large efforts into compensating atmospheric influences. However, even in the era of nano- and microsatellites, space observatories are intrinsically expensive and bear operational limitations: development times are long, updates or corrections of the instrumentation are usually not possible after launch, operating material such as cryogenic coolant fluids cannot be refilled or replaced (see the Herschel Space Observatory). Furthermore, comparatively conservative approaches towards new technologies are used to minimize risks of expensive failure. Ground-based and airborne telescopes, on the other hand, still suffer from fundamental limitations imposed by the atmosphere at certain wavelengths.

On the other hand, technological advances have made balloon-borne telescopes more attractive over the last couple of years. These particularly include more reliable balloons and the opening of long-duration flight routes, allowing flight durations of 30 to 40 days on "conventional" long duration routes and promising 50 to 100 days on ultra long duration routes. Consequently, the last couple of years saw an increase of balloon-astronomy initiatives aiming at more regularly flying missions rather than the more common "experiment"-type of flights. Noteworthy recent examples are the U.S./Canadian Superpressure Balloon-borne Imaging Telescope (SuperBIT) [1], the JPL-lead Astrophysics Stratospheric Telescope for High Spectral Resolution Observations at Submillimeter-wavelengths (ASTHROS) [2], and the U.S.-lead Balloon-borne Large Aperture Submillimeter Telescope (BLAST) [3] along with its successors.

The goal of ESBO is to further lower the entry barrier to balloon-based observations by providing an operating institution that offers observing time and instrument space on balloon-based telescopes.

## 1.2 ESBO project history and future

From 2016 to 2018, the H2020-funded project ORISON (innOvative Research Infrastructure based on Stratospheric balloONs) assessed interest and scientific needs within the (mostly European) astronomical community with regard to balloon-based research infrastructures and studied the general feasibility of a balloon-based observatory [4].

The plans for ESBO pick up from the positive conclusions of ORISON and aim at creating an observatory institution based on the following cornerstones:

- Provision of instrument flight opportunities as well as open observation time access to the scientific community,
- Operation of regularly flying balloon telescopes and provision of related services by an operating institution,
- Provision of the regular opportunity to refill consumables, upgrade and/or exchange instruments in between flights,
- Maximum reuse of platform hardware in between flights to ensure efficient and fast turnaround in between flights, which includes safe recovery of hardware.

More details on the envisioned ESBO infrastructure and its scope can be found in section 2 of this paper.

The ongoing ESBO *Design Study* (ESBO *DS*) represents the second step towards ESBO. Under ESBO *DS*, the full infrastructure, with particular technical focus on far infrared (FIR) observational capabilities, is being conceptually designed. In addition, a prototype UV/visible flight system (STUDIO) is being developed and built to test some of the key technologies identified.

The concept for the FIR capabilities is presented in section 3 of this paper. Section 4 describes the STUDIO prototype in detail. Finally, sections 5 and 6 shortly outline selected aspects of ESBO *DS* that are also relevant for space missions and the critical technologies for a full exploitation of the potential that stratospheric balloons offer, particularly for FIR astronomy.

# 2. The ESBO concept & plans

## 2.1 Operational concept

The long-term goal for ESBO is to establish an operating organization that conducts regular flights (1+ per year) of balloon-borne telescopes. At least the gondola and the respective telescope(s) are thereby foreseen to be provided by the operating organization as well. With regard to instrument provision, we regard it as desirable to foresee both "facility instruments", with open access to observation time, as well as flight opportunities for "PI instruments". This approach should make ESBO a service provider for both instrument builders and observers.

These considerations require that the ESBO infrastructure does not only include flight systems, but that it also offers tools and procedures for the instrument preparation, proposal, observations, and data reduction phases

While the current concept foresees that an ESBO institution provides flights of balloon-borne platforms, it also foresees that the operation of launches and of the balloon itself during flight be procured from currently existing and experienced launch and flight providers.

*2.2 General tools for balloon-based astronomy*

As part of the observatory approach, ESBO has started and will continue to develop several tools primarily aimed at supporting instrument developers. These tools are and will be made available to the scientific ballooning community in general as well. Particular activities that already have started and that we plan to maintain and extend in the future are:

- Development and provision of ECSS*-adapted test procedures for balloon-qualified components, parts, and materials;
- Preparation and provision of a database of pre-flown and qualified components, parts, and materials;
- Support of the development and preparation of community-available tools relevant for balloon-astronomy, such as schedulers, on-board software, etc.

*2.3 Complementarity to space missions*

Balloons can cater a partly overlaying, partly different parameter space of mission requirements compared to space missions.

On the one hand, individual flights are considerably shorter than space missions. Depending on the exact wavelength, some atmospheric absorption may still exist. Particularly interesting for infrared observations, cooling of large structures, e.g. telescope mirrors, to cryogenic temperatures is much more difficult than in space.

On the other hand, the comparatively much lower mission costs in combination with the shorter flight durations (and re-flights to obtain more observation time) allow a less risk-adverse approach. This, in turn, allows shorter development times, connected with the possibility to fly more up-to-date instrumentation. Regular returns of instruments furthermore allow their update and re-arrangement, including, of particular interest for infrared observations, the possibility to refill cryogens. Balloons furthermore come with less solid restrictions to the geometrical size of payloads, which are basically suspended in free air underneath the balloon.

These aspects suggest to use balloon-based observatories not as much to obtain extremely high sensitivity, but rather to obtain high spatial resolution, to carry out surveys, to cater rapidly developing or versatile instruments, and to address instrumental and observational gaps that for one reason or the other are not covered by space-based observatories (such as heterodyne instrumentation for the FIR, for example). Given the lower required investments and smaller timescales, they also tend to allow more flexibility and adjustability for specific scientific interests.

Taking the FIR as an example, currently no space-based observatory exists for this wavelength domain, and new space-based observatories would be available in 2029/2030 the earliest. Leading the possible schedule is the Space Infrared Telescope for Cosmology and Astrophysics (SPICA), for which the decision on its selection for ESA's M5 mission is foreseen to be taken in 2021. The second large FIR space mission under consideration is the Origins Space Telescope (OST), currently submitted for review under the 2020 U.S. Astrophysics Decadal Review. The launch of OST is not planned for earlier than 2035. SPICA thereby focuses clearly on sensitivity and will not have higher spatial resolution than the airborne Stratospheric Observatory for Infrared Astronomy (SOFIA). OST would carry a large mirror, but neither of the two concepts includes heterodyne, i.e. extremely high spectral resolution instrumentation.[†]

The following chapter describes this gap of capabilities in the FIR and how balloon-based observations can fill it in more detail.

## 3. Concept for a balloon-based FIR observatory

As the FIR range is not accessible from the ground, the FIR astronomy community only slowly started to develop (and is still developing) thanks to a series of space-based and airborne observatories. Most of them, however, have been spacecraft with a limited lifetime or instruments with limited spectral range or resolution. As such, after the end of the Herschel mission, the community is currently left with only one active FIR observatory, SOFIA, and sporadically flying balloon missions (such as BLAST [6], the Stratospheric Terahertz Observatory (STO) [7], or the Polarized Instrument for the Long-wavelength Observation of the Tenuous ISM (PILOT) [8], see also Table 1).

---

* European Cooperation for Space Standardization

† The highest resolution instrument currently foreseen in these concepts is the Origins Survey Spectrometer (OSS) for OST, with a spectral resolution of up to R~3*10^6 [5], still considerably below the resolution at the order of $10^7$ provided by heterodyne instruments. OSS would furthermore only provide this resolution with one pixel, making it unsuitable for efficient mapping.

Table 1. Active far infrared observatories.

| Mission | Wavelength coverage | Effective aperture diam. |
|---------|---------------------|--------------------------|
| SOFIA[‡] | 0.36 – 612 μm | 2.5 m |
| BLAST [10] | 240, 350, 500 μm | 2.5 m |
| PILOT [6] | 240 & 550 μm | 1 m |
| STO [5] | 158, 205 μm | 0.8 m |

Astronomers, especially astrochemists, are still waiting for new FIR telescopes. It is thus the time to plan the next mission that will cover the gap in the FIR sky between JWST's upper limit (28.5 μm) and ALMA's lower limit (316 μm, Band 10).

Next steps of FIR science will further investigate the origins of water on planets in our and distant solar systems, study mechanisms and details of star and planet formation by investigating chemical evolution and cooling processes throughout the universe, and further investigate the Interstellar Medium, its interaction with stellar environments, and its energy cycle, by observations of dust and gas (as expressed e.g. by the European Far-Infrared Space Roadmap [9]).

Taking these scientific steps forward requires telescopes with better angular resolution, more observational capacity (in terms of spectral coverage and observation time), and higher sensitivity. The latter need would be addressed by SPICA (2.5 m telescope diameter) [10]. The first need, however, can only be achieved by larger telescopes or interferometric observations. In addition, large telescope apertures are required in order to overcome the so-called confusion limit, the notion that strong signals from dust and gas in the foreground make small astronomical targets in their background indistinguishable. Logistical and technical challenges, however, make space-based telescopes with large apertures, as required for high angular resolution observations, extremely difficult. Balloon-borne telescopes do not face many of these challenges and thus are particularly well suited to address the first two needs, while offering the possibility to regularly use the most up-to-date instrumentation. Such a concept has already been proven successful by SOFIA.

---

[‡] Current coverage in Observing Cycle 9, wavelength coverage partly interrupted by atmospheric extinction [11].

### 3.1 Scientific motivation
#### 3.1.1 Discrete sources

Particularly the advancement of research concerning star and planet formation and astrochemistry, but also solar system science, relies upon further information from FIR observations. Outstanding topics in these areas include the further study of cold dust and ices, that of light hydrides, and the study of the distribution of molecules in general, be it in atmospheres of solar system objects, the Milky Way, or other galaxies. In the following, we will highlight a few of the prominent science case whose study a large aperture balloon observatory would enable.

Ice features in the FIR

Dust, for years, annoyed astronomers by covering their favourite stars as well as the birth places of those stars and their planetary systems. With the development of infrared observatories, however, cold dust and ices became a hot topic, allowing important insight into the process of star and planet formation and into the migration process of water through evolving planetary systems. So far, mainly features in the near- and mid-infrared have been used to detect and characterize molecular ices in dark clouds and protoplanetary disks. As their FIR emission features are attributed to intermolecular vibration modes, however, observing them in the FIR makes it furthermore possible to determine the structure and transitions between phases of the observed medium (e.g. amorphous vs. crystalline). In particular, the FIR band positions and widths are, in addition to the abundance of the emitting species, sensitive to the grain geometry and size distribution, the environment temperature and density structure. Combined with modelling of protoplanetary disk emissions, the analysis of the FIR features thus allows to infer the abundance and location of ices within the disk, making it, with sufficient data, possible to constrain the location of the snow line (the distance from a star/protostar where it is cold enough for volatiles to condense into ice) [12].

So far, only water ice features have been detected in the FIR in a few disks, while the band strengths of other ices are thought to be not strong enough to have been detected. A sensitive balloon observatory offering medium spectral resolution observations at the wavelengths of these ices would help to study them in many targets across the Galaxy (for band locations and band strengths of prominent molecular ices from laboratory measurements see Giuliano et al. [14]).

Light hydrides

Light hydrides, on the other hand, belong to the first molecules to form in atomic gas and are thus at the starting point of astrochemistry and the building blocks of larger molecules [14]. Their study allows fundamental insight into the first building steps towards interstellar molecules. As their chemical formation process only involves a few steps, the interpretation of their abundances is comparably straightforward and they can provide key information about their environments, including on dynamical processes (shocks, turbulence, large scale winds), cosmic ray ionization rate, and presence of molecular hydrogen [14].

The observation of light hydrides thus promises to provide a valuable tool to understand planet and star formation, and, through the measurement of isotopic ratios, also to understand the origin of volatiles in our own solar system. While the idea behind the study of light hydrides sounds relatively simple, their observations are much more complex. They require very high spectral resolution mostly in wavelength ranges that are not accessible at lower altitudes. ALMA and NOEMA are already enabling highly sensitive and spatially highly resolved observations of light hydrides in distant galaxies, for which the ground state transition lines are sufficiently redshifted to fall into the sub-mm/mm spectral regions. For observations in the interstellar medium in our own Galaxy, or neighbouring galaxies, however, the lines have to be observed at (or close to) their original wavelengths in the FIR. A sensitive balloon observatory offering high spectral resolution observations at the wavelengths of light hydrides ground states would thus allow their study in many targets across our Galaxy.

### Solar System atmospheres

In the solar system context, balloon-based observations would allow e.g. studies of the vertical distributions of molecules in atmospheres of planets, their satellites, or in comae. These observations require high spectral resolution to determine the shape of absorption lines. While many molecules could theoretically be observed with SOFIA, telluric absorption lines are still considerably pressure broadened in the remaining atmosphere. At 30 to 40 km altitude, the line width of telluric absorption lines is narrow enough to allow to distinguish between the telluric lines and Doppler shifted absorption features on solar system objects.

### 3.1.2 Surveys

The FIR spectral range is generally lagging behind with regard to high-resolution large-scale maps of our Galaxy, as compared to the adjacent sub-mm and mid infrared spectral regions. The only complete continuum sky survey at 100 µm existing today, for example, is the one taken by the Infrared Astronomical Satellite (IRAS) with a resolution of 1.5 arcminutes. The situation is similar with regard to dedicated spectral line surveys. Particularly the 157.7 µm (CII) line and the 63.18 µm (OI) line may radiate up to several percent of the entire Galaxy's energy output, and Herschel and SOFIA have been able to study these important lines a bit already. However, very little is still known about their spatial distribution across our Galaxy. Without capacities in the FIR, this situation will not change, as Figure 1 indicates. Mapping out these and other important FIR spectral lines (such as the 128 µm HD line) across a major fraction of our own Galaxy as well as of other galaxies with a high signal to noise ratio will boost our understanding of the chemical evolution of our Galaxy and of galactic evolution in general.

A balloon observatory can achieve about 1000 hours of observing time during a 6-weeks mission with one instrument attached. Such a set-up is very much suited for executing large surveys.

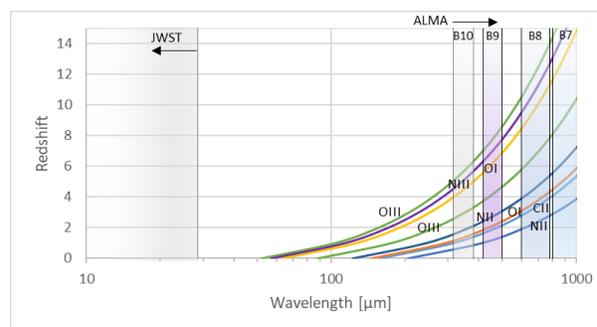

Fig. 1. Spectral locations of major atomic lines, including the OI and CII cooling lines at different redshifts. At the left and right ends of the figure, the spectral ranges covered by JWST and the shorter-wavelength ALMA bands (10, 9, 8, and 7) are marked. As the figure indicates, in our own Galaxy these lines can only be observed in the far infrared.

### 3.2 Potential implementation/technical concept

For the potential technical implementation of a balloon-based 5 m aperture FIR telescope, we consider, in contrast to an earlier approach, a slightly modified Cassegrain layout employing a primary mirror carried out as a carbon fibre reinforced polymer (CFRP) sandwich structure. For a mirror strictly optimized for the far infrared, we estimate that an areal density of 18 kg/m$^2$ is achievable, leading to a total mirror mass of approx. 350 kg (even lower areal masses of FIR CFRP mirrors have been achieved for smaller mirrors in the past, such as with the first BLAST mirror (10.2 kg/m$^2$ at 2 m diam.) [15] or the CFRP mirror designed for Herschel (11.4 kg/m$^2$ at 3.5 m diam.) [16]).

In order to ease balancing and the exchange of instruments, we consider using an "adapted" Coudé focus reached via two relay mirrors on the telescope tilt axis in between the primary and secondary mirrors.

A second challenge besides the telescope size is the power consumption. A heterodyne instrument providing very high spectral resolution, particularly when considering an extended array, has comparatively high power requirements. The constant sunlight conditions of a polar flight are advantageous in this regard, however. Using a conventional open-loop cooling approach for the instrument to not further increase the power requirement (a closed-loop cryo-cooler would increase the average required power by several kW), we estimate a required solar array area of 35 m².

Overall, we estimate a total mass of the balloon payload, including ballast for 40-day flights, at approx. 4800 kg. While this would not allow flights at 40 km, it may well be feasible to modify current balloons to carry this mass for flights at 30 km altitude, which is sufficient for observational purposes.

### 3.3 Potential performance

While the main strength of a large balloon-based FIR telescope would be in the angular resolution rather than in sensitivity, the good observation conditions are also reflected in the expected sensitivities.

For a state-of-the-art high spectral resolution heterodyne instrument, without assuming any improvements on the instrument, the sensitivity would improve greatly on the conceptual ESBO FIR platform, as Figure 2 indicates for some of the prominent FIR atomic and molecular lines.

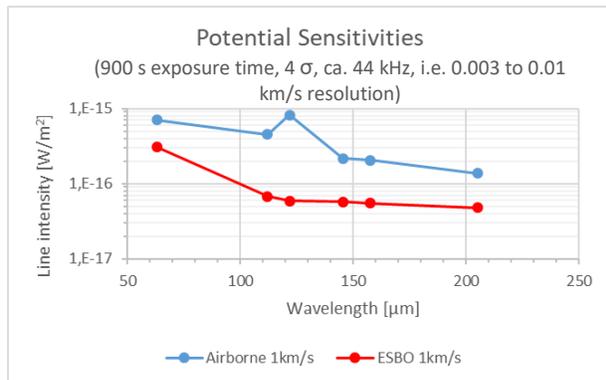

Fig. 2. Estimated minimum detectable line fluxes for single point observations of prominent atomic & molecular lines (OI, HD, NII, OI, CII, NII) for a high spectral resolution heterodyne instrument at 30 km altitude. Values in between data points are interpolated.

In addition, many further spectral regions that are not observable at 14 km become accessible at 30 km. This is particularly well illustrated by the case of light hydrides, where many lines become observable only at balloon altitudes. Table 2 lists a selection of those lines, along with sensitivity estimates for high spectral resolution (44 kHz) observations.

Table 2. Estimated minimum detectable line fluxes for high spectral resolution (ca. 44 kHz) of selected lines of light hydrides. Sensitivities are 4 σ, for 900 s exposure time and assumed line widths of 1 km/s.

| Species | Wavelength | Line sensitivity [W/m²] | |
| | | 14 km | 30 km |
| --- | --- | --- | --- |
| $H_3O^+$ | 181.05 µm | $8.95 \cdot 10^{-16}$ | $4.68 \cdot 10^{-17}$ |
| $H_3O^+$ | 100.87 µm | - | $7.97 \cdot 10^{-17}$ |
| $H_3O^+$ | 100.58 µm | $5.82 \cdot 10^{-16}$ | $7.61 \cdot 10^{-17}$ |
| $CH^+$ | 179.62 µm | - | $4.90 \cdot 10^{-17}$ |
| $CH^+$ | 90.03 µm | - | $1.46 \cdot 10^{-16}$ |
| $CH^+$ | 72.14 µm | $5.59 \cdot 10^{-16}$ | $2.79 \cdot 10^{-16}$ |
| HF | 121.70 µm | - | $8.14 \cdot 10^{-17}$ |
| HF | 81.22 µm | - | $8.30 \cdot 10^{-16}$ |

### 3.4 Flight options

Due to the decrease of the efficiency of Rayleigh scattering with increasing wavelength, the FIR daytime sky at balloon flight altitudes is not considerably brighter than the FIR nighttime sky. Consequently, from an observational point of view, FIR observations can be carried out during any time of day. This also opens up the theoretical possibility to use any available balloon flight route. Detector and cooling systems for FIR instruments tend to be power hungry, however, and particularly for large and voluminous instruments, thermal stability is not easily achieved in a thermally changing environment. Additionally, Zero Pressure Balloons (ZPBs), which are currently the only balloons to offer the required payload capacity for a 5 m-aperture FIR telescope, can only be operated for extended times under relatively stable thermal conditions.

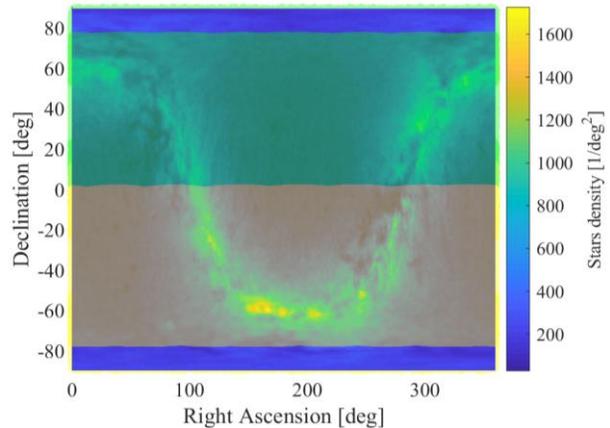

Fig. 3. Accessible sky using a combination of Northern and Southern circumpolar flights. Green shaded area shows accessibility from Svalbard, yellow shaded area shows accessibility from McMurdo (each for a 40 days flight).

Due to these reasons, polar flight routes during respective summer are advantageous for FIR observations. The Southern circumpolar route is well developed, with regular launches from McMurdo and 30

 

to 40 days of flight using "conventional" ZPBs. The Northern circumpolar route is currently not in regular use for full circumpolar flights. It might, however, be worthwhile to pursue using this route again. Using both the Southern and Northern circumpolar routes (e.g. launching from Svalbard, Norway in the North), and when allowing a minimum elevation of 12° for observations, almost the entire sky would be accessible with two flights, as Figure 3 shows.

## 4. STUDIO – the UV/visible prototype mission

A major part of ESBO *DS* is the development and construction of the STUDIO prototype platform. The STUDIO prototype comprises a versatile, modular gondola for astronomical applications, carrying a 0.5 m aperture telescope suitable for observations in UV to near infrared (NIR) wavelength ranges. As a main scientific instrument, it will carry an advanced photon-counting, imaging microchannel plate (MCP) detector on the first flights. STUDIO will thereby serve as both a technology demonstrator and testbed, as well as a platform for astronomical observations.

### 4.1 Scientific motivation

Astronomical observations in the UV at wavelengths below ~320 nm are not possible from the ground because of atmospheric extinction by ozone. However, at an altitude of 37 km, observations are feasible down to ~200 nm. Therefore, the balloon-borne STUDIO enables UV observations that would otherwise only be possible with space-based telescopes. STUDIO comprises two simultaneously operating imaging instruments as a UV and a visible/NIR channel. Two cases motivate the UV scientific part of STUDIO.

### 4.1.1 Search for variable hot compact stars

Hot and compact stars are the rather short-lived end stages of stellar evolution. They comprise the hottest white dwarfs (WDs) and hot subdwarfs. A significant fraction of them show light variations with periods ranging from seconds to hours. Among them are diverse types of pulsators, which are important to improve asteroseismic models. Others are members of ultracompact binaries (e.g., WD+WD pairs) and are strong sources of gravitational wave radiation and crucial calibrators for the future space mission eLISA. Hot compact stars have so far been studied predominantly at high Galactic latitudes. Due to their very blue colours they stick out in old stellar populations like the Galactic halo. However, the density of stars at high Galactic latitudes is rather small and those objects therefore very rare. Due to the 1000-times higher stellar density, the Galactic disc should contain many more of those objects. Searches in the Galactic plane are desirable, but the identification of these faint stars is hampered by the dense, crowded fields. But not so in the UV band. The hot stars are much easier to detect there, because their emitted flux is increasing towards the UV, while the flux of most other stars decreases because of their lower temperatures. Surveying the Galactic plane with a UV imaging telescope will uncover many new variable hot stars.

### 4.1.2 Detection of flares from cool dwarf stars

Red dwarf stars (spectral type M) are hydrogen-burning main sequence stars like our Sun, but less massive, cooler, and less luminous. A large fraction of the stars in our Milky Way belongs to this group. They emit most of their radiation in the visible and NIR wavelength regions. Their UV and X-ray emission, despite being energetically a minor contribution to the overall radiation budget, ionizes material surrounding the stars and is, therefore, of central interest for the evolution of planets and other circumstellar matter. This high-energy emission is highly dynamic. One characteristic phenomenon are flares that are stochastic brightness outbursts resulting from reconfigurations of the magnetic field. During such flares, these normally faint stars become much brighter for the duration of minutes. A strong emission line of ionized magnesium (Mg II) at 280 nm, covered by the STUDIO instrument, can carry up to 50% of the near-UV flux during flares. Up to now, no systematic UV monitoring of "flare stars" exists. Consequently, the flare occurrence rate is unknown as well as the flare energy number distribution. Particularly interesting for the study of the physics of flares is their multi-wavelength behaviour (time lags, relative energy in different bands). However, only a few simultaneous UV and optical observations exist. STUDIO enables such observations by continued monitoring of prominent objects.

### 4.2 System architecture

The STUDIO platform is logically divided into the gondola/bus part and a separate payload part as illustrated in Figure 3. The communication is automatically switched between line-of-sight (direct radio link) and beyond-line-of-sight (Iridium) by the bus system, providing a transparent tunnel for the payload communication. Pointing is commanded by the payload computer on board, but due to the complexity of the pointing system, a separate monitoring and control station is set up on ground. This overall split-up allows a reasonably easy exchange of instruments.

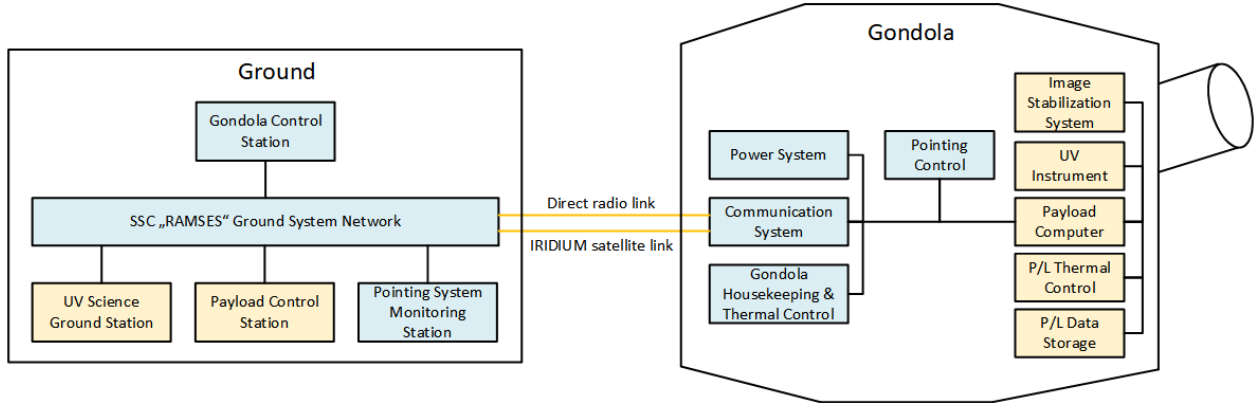

Fig. 3. STUDIO system architecture. Bus components in blue, payload components in yellow.

### 4.3 Payload

#### 4.3.1 MCP detector

The main science instrument of STUDIO is an imaging and photon counting MCP UV detector, developed and built by the Institut für Astronomie und Astrophysik Tübingen (IAAT). It is a successor of the Echelle-spectrometer detector whose development and flights for the ORFEUS missions are part of the space heritage of IAAT [17]. The detector is sealed by a UV-transparent window (fused silica/MgF$_2$/LiF) to maintain an ultra-high vacuum, which is needed for a proper photocathode operation. The photocathode converts a photon into a photoelectron with a conversion rate depending on the photocathode material and the wavelength of the incoming photon. A quantum-efficiency of up to 70% at 230 nm seems achievable for GaN [18]. In the MCP stack, a high voltage accelerates the incident photoelectron, resulting in a charge cloud at the bottom side of the MCP stack. The position information of the incident photoelectron is preserved in the MCPs as the centre of charge of the electron cloud. The electron cloud is accelerated towards a cross strip anode (64 by 64 strips). The charge signals of the 128 anode-strips are amplified by a BEETLE chip [19]. The output of the chip is converted into digital signals and sent to the processing unit, a Virtex-5 FPGA. The implemented centroiding algorithm will calculate the centre of charge for each event with an accuracy of 1/32 of the distance between two anode strips, resulting in an image resolution of about 2000 x 2000 pixels.

#### 4.3.2 Telescope & optical system

STUDIO's optical payload is composed of a 50 cm aperture modified Dall Kirkham telescope to which will mounted the Telescope Instruments Platform (TIP).

The telescope's secondary mirror mechanism includes three actuators remotely operable for focusing the mirror with a resolution of ~ 3 μm.

The TIP includes two principal instruments:

- An advanced photon photon-counting, imaging MCP detector (see section 4.3.1) that will perform observations in the UV band [180 nm-330 nm]
- A commercial visible light camera used mainly as the tracking sensor in a closed closed-loop fine image stabilization system, but will also serve as a secondary scientific instrument to cover for observation in the VIS band [350 nm-1000 nm]

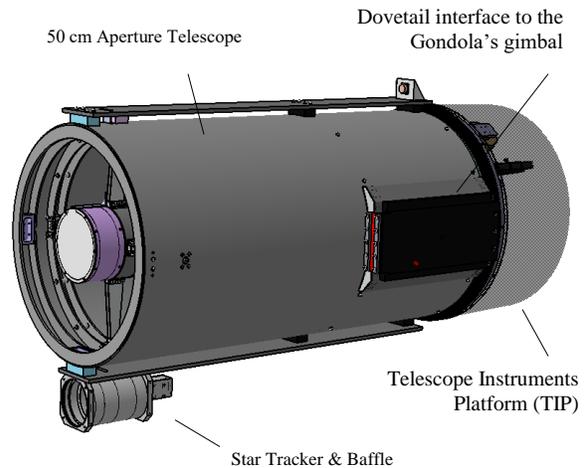

Fig. 4. Telescope + TIP assembly

A filter wheel is placed in front of each instrument. Five standard Sloan filters and one empty slot are foreseen.

Additionally, the TIP houses a fast steering mirror mounted on a commercial Tip/Tilt Platform as part of the image stabilization system to achieve 1 arcsec pointing stability (see also section 4.3.3).

A dichroic mirror divides the light beam into two channels by reflecting the UV light [180 nm-330 nm] and transmitting the visible light [330 nm-1000 nm]. The two separated beams are then redirected to the corresponding detector through highly reflecting mirrors.

All instruments are mounted on a stiff but lightweight CFRP – aluminium honeycomb sandwich plate.

In order to maintain optical alignment of the components over the expected temperature range, the mechanical mounts are designed to be thermally self-compensating. The instruments are kept from overheating using a passive control system. Temperature sensors and contingency heaters are foreseen for all critical systems and will to keep them within their operational temperature range.

The TIP is closed using a CFRP enclosure and sealed to keep the optics from possible contamination. Contamination protection is supported by a small overpressure (~15 mbar) with dry nitrogen in the optics compartment during ascent.

### 4.3.3 Image Stabilization System (ISS)

The ISS will compensate the remaining image jitter after the gondola pointing system of up to ± 40 arcsec (peak-to-peak) with frequencies of up to 40 Hz in order to stabilise the image on the focal plane to a maximum residual jitter amplitude of 1 arcsec (peak-to-peak). It is designed to maintain the image stability over 300 seconds (90% of the time in which the remaining input jitter is within the limits of the interval of frequencies). For this purpose, the ISS employs a visible light tracking camera with a baseline exposure time of 12.5 ms. Given the high required read-out rate, the system is compatible with different sub-frame sizes ranging between 80 x 80 arcsec and 200 x 200 arcsec in the sky. The ISS is designed to work on guide stars with a V-band magnitude down to 9.

The point spread function will be larger than the required maximum jitter amplitude (80% encircled energy within ca. 1.5 arcsec diameter) so that centroid tracking of the point spread function of the guide star will be used.

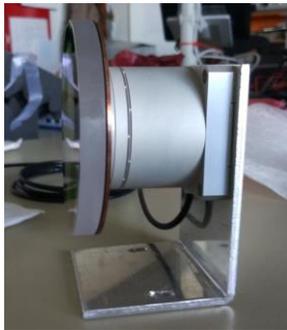

Fig. 5. Provisional mirror with a brass disc mounted on the Tip/Tilt platform for laboratory tests.

Currently, the software for the ISS is under development following the specific architecture defined for the STUDIO payload on-board (see section 4.3.5).

For this purpose, we acquired a piezo Tip/Tilt platform S-340.ASD and a multichannel digital piezo controller E-727.3SDP. For the test setup, we use a fused silica mirror of 100 mm diameter coated with aluminium with a brass disc glued to its back in order to compensate the mass difference to the flight model. Both together are mounted on the platform. We are making mechanical tests in the laboratory for observing the micro-movements and improving the data acquisition of the centroid with an openCV application, which allows showing the image in real time. For this, we are using a CCD camera ZWO ASI120mm-S aligned to the mirror.

Figure 5 shows the entire platform mounted on an aluminium mount.

### 4.3.4 Payload electronics

During the development of the STUDIO payload electronics, a focus was set on the use of readily available commercial off-the-shelf (COTS) hardware, leading to significant reductions in cost and development time.

The use of an unmodified COTS payload computer and commercial SSDs was enabled by enclosing these components in a pressure housing. A commercially available camera was qualified for operation under stratospheric conditions, eliminating the need for a pressure housing on the optical bench.

Modifications were made to COTS filter wheels and the control electronics for the telescope's focussing mechanism in order to harden them against the harsh conditions in the stratosphere (particularly temperature and vacuum). Subsequent qualification tests ensured the effectiveness of these enhancements.

However, the use of COTS hardware or affordable aerospace grade hardware was not possible for all functions. Therefore, a device to control and power the payload electronics and a system to measure various environmental variables were developed and qualified at the University of Stuttgart.

The use of COTS components, some even unmodified, and the in-house development of components clearly highlighted the need of standardized testing. Therefore, qualification procedures are being developed and will be improved to include experience gained throughout the project. This will also facilitate and accelerate the development of future scientific instruments, since working groups are spared the development of their own qualification procedures.

### 4.3.5 Spacecraft-derived on-board software

The on-board software for STUDIO is based on the Flight Software Framework that was developed for the Flying Laptop satellite and has a centralized architecture. It is a component-based software with a core framework that interfaces with different devices and handles inter-process communications between these components.

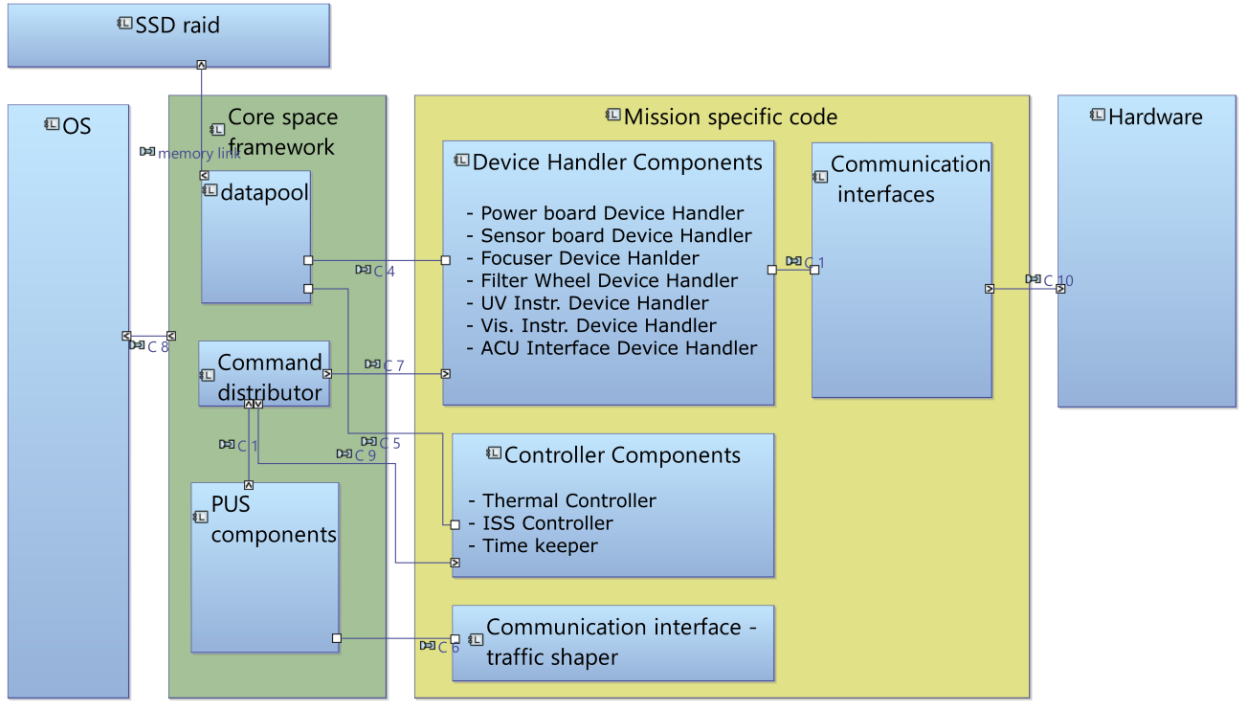

Fig. 6. STUDIO on-board software architecture. (OS: Operating System; ACU: Attitude Control Unit)

The framework is platform and operating system independent. ESBO will use Linux as operating system on-board on an industrial single-board computer, as there are no hard real-time requirements for the prototype mission.

Three different component types are defined in the software framework. These include device handlers, controllers, and communication services. Each of these types has a template that facilitates implementing a device software and defines the interfaces to the framework and other components. In addition, as the core framework is mainly developed for space missions, ESBO uses ECSS Packet Utilization Standard (PUS) for standard telecommands and telemetry communication with the payload.

Figure 6 shows the architecture of the ESBO on-board software. Mission specific code includes several devices and controllers, all exchanging data with the global datapool and command distributor. On the other side, device handlers are flexible as they are not implemented for a specific lower level interface. They can communicate with any communication interface as long as it is wrapped in the required software interface. PUS services have recently been added to the core framework to increase reusability of the framework for future missions.

### 4.3.6 Operation of payload software

The core software framework provides flexible mode-based operations onboard. Automatic operations can be carried out in different operational modes, just by setting the mode of the device. Devices and controllers can be grouped together in subsystems, and if one fails to operate normally, the whole subsystem configuration can automatically respond. While STUDIO does not include redundancies at this point, this feature will be used in future missions to add more automation and increase reliability and robustness.

Apart from the mode management onboard, the on-board software uses pre-loaded observation commands (stored in the on-board mission schedule) that are released to the devices at specific time lines. These commands can be loaded from a file when the mission starts. The commands for each observation field will be grouped, and each group can be shifted in time or disabled later via ground control. In addition to the pre-loaded commands, STUDIO also allows edition of the mission timeline via the permanent link to the ground station.

### 4.4 Gondola

The gondola/bus supports the payload with all essential service systems. The mechanical structure is designed from COTS aluminium tube structures in combination with custom-made tubing structures and holds the payload and all the subsystems. The idea is to have a flexible design that easily can be modified for different payload requirements and sizes and that can also conveniently be disassembled in the field for easy recovery. The structure is covered with white fabric for sun protection that can easily be removed during testing. The telescope and the pointing system is mounted directly in a stiff inner gimbal-like structure, as it can be seen in Figure 7.

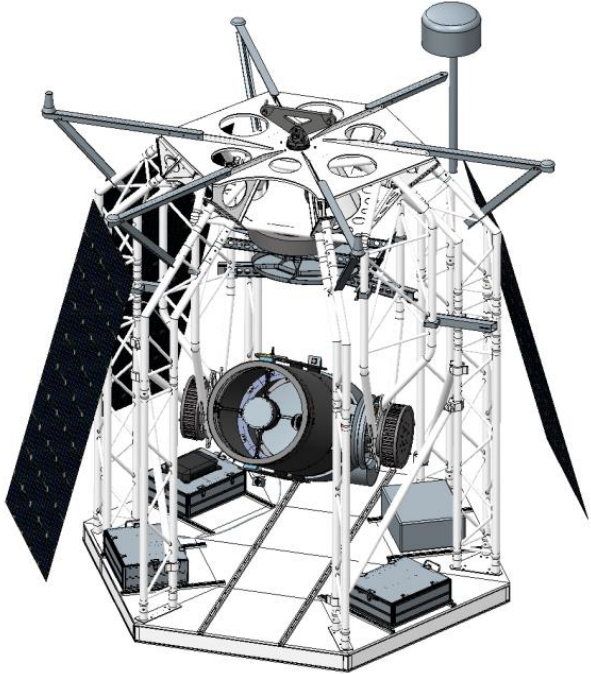

Fig. 7. STUDIO gondola. One solar panel is not shown for better visibility of the payload. Most electronics and service systems are located on the gondola floor.

The pointing system is a further development of the pointing system developed for the PoGo balloon missions [20, 21]. It is equipped with an elevation motor directly operating on the telescope and azimuth motors and a fly wheel for turning the complete gondola including the telescope. The pointing system uses multiple sensors, but three differential GPS antennas and a star tracker are the most essential sensors.

The other major systems are the power system and the communication system that are based on previous SSC designs from sounding rocket and stratospheric balloon missions.

The power system comprises the solar panels, batteries and electronic system for power management. It supplies all the systems with power in the gondola including instrument and pointing system. The batteries and electronics are placed in COTS aluminium boxes for thermal and mechanical protection The system can easily be scaled according to the mission requirements by increasing or decreasing the number of power boxes. In the STUDIO mission, the system is scaled with batteries for covering the nighttime observations and solar panels for daytime charging of batteries. A qualification process for qualifying COTS solar panels is ongoing and, if successful, it will bring down the cost significantly and reduce delivery time by using COTS panels.

The system also comprises electronics and software for controlling the power distribution, monitoring of voltages and currents and switching on and off for the different power outlets. It also comprises housekeeping functions such as temperature, pressure, and GPS signal that is distributed to the payload.

The communication system that is used for the gondola (for control of the balloon a separate communication system is implemented for safety reasons) has two separate communication links. For close range communication, when the balloon is within a couple of hundred kilometres from the ground station, a high bandwidth link is used. The range can be extended by using extra ground stations that are placed on the predicted trajectory. For over the horizon or global communication an Iridium-NEXT system will be used with lower bandwidth. This second generation Iridium system gives higher bandwidth and also smaller antennas compared to the previous Iridium systems used in ballooning.

*4.5 Flight plans*

The first flight of STUDIO is planned for late summer of 2021 or 2022 over Esrange, Sweden. The choice of time allows us to take advantage of the seasonal "turnaround conditions" of the stratospheric winds, which enable payloads to remain practically above the launch site for about 40 h. Combined with the available infrastructure at Esrange Space Center and the comparatively to other launch sites easy logistics, this setup is very suitable for a first test and science flight. Choosing the turnaround window in August rather than in April / May furthermore allows us to take advantage of the seasonally lower ozone content in the Northern hemisphere, which constitutes the main atmospheric absorbent for UV radiation (the total ozone column over Northern Sweden differs between around 400 Dobson units in late April and around 300 Dobson units in late August).

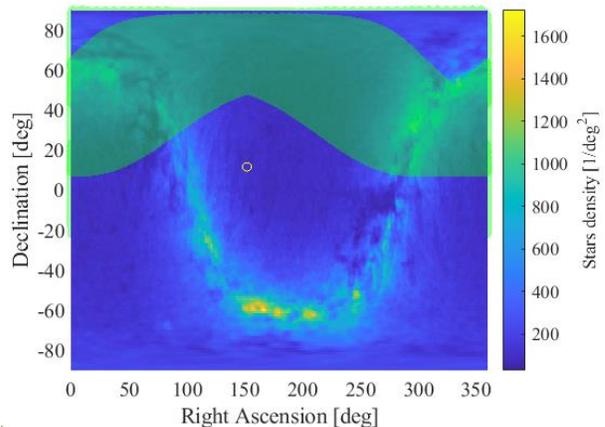

Fig. 8. Accessible area of the sky during a summer turnaround conditions flight over Kiruna

Figure 8 shows the accessible area of the sky during the planned flight. We aim at observing objects in the Galactic plane, to which this flight option provides us with sufficient access.

Table 3. Summary of flight details for the first STUDIO flight.

| Duration | 40 h |
|---|---|
| Altitude | 37 km |
| Location | Esrange, Sweden |
| Time | August 2021/2022 |
| Gondola mass | 743 kg |
| Balloon size | 600,000 m³ |

In order to lift the 743 kg gondola to 37 km, which are required to reach the necessary atmospheric transmission in the UV (see also Figure 9), we are planning to fly with a 600,000 m³ zero pressure helium balloon.

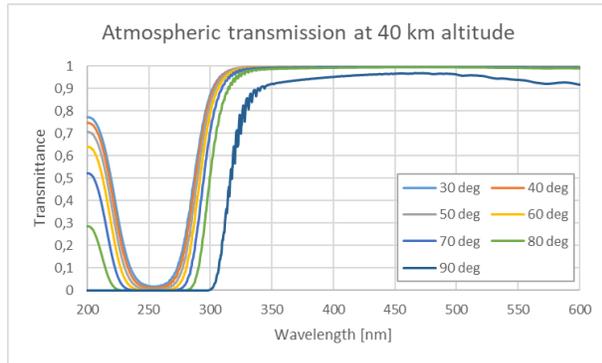

Fig. 9. Atmospheric transmission in the UV at 40 km altitude for different zenith angles (MODTRAN simulations). Simulations show that the transmission at 200 nm increases by more than a factor of 2.5 between 30 and 40 km altitude, eventually resulting in a required flight altitude of 37 km.

## 5. Technologies extensible to space missions

STUDIO does not only serve as a prototype mission and testbed for further balloon astronomy platforms under the ESBO initiative, but also features development and test activities that will be used in future space missions. The two most prominent aspects, the MCP detector and the flight software, are shortly described in more detail in the following.

### 5.1 MCP detector development for space missions

MCP detectors developed and designed at IAAT shall be utilized in space instruments performing astronomical observations in the UV. TINI (Tübingen IAA Nebular Investigator) is a far-UV spectroscopic imaging instrument currently being developed at the Indian Institute of Astrophysics (IAA) at Bangalore in collaboration with IAAT. The Census of WHIM

Accretion Feedback Explorer (CAFE) is a multi-channel far-UV imager being planned by the Purple Mountain Observatory at Nanjing in collaboration with IAAT. Scientific requirements of both missions are such that MCP detectors with high quantum efficiency are needed, which is offered by our detectors for which we produce GaN photocathodes [22]. Our detectors are characterized by long lifetime, high dynamics, small volume, low mass, and low power consumption.

### 5.2 Spin-off and spin-in of flight software

The ESBO goal as an infrastructure is to provide the means to accommodate different instruments in a short time-span and fast and economical access to scientific results. This flexibility of integration is also required for on-board software. While distributed computing would also be an option for stand-alone modules, it constraints complex scientific operations that would require exchange of information among several components and use of centralized resources.

STUDIO (and also future ESBO missions) uses a flexible component-based framework with clear interfaces as a core software. This framework is a direct spin-off from the Flying Laptop satellite mission. This framework facilitates the goal of developing a device handling software and integrating the device with the centralized on-board software. The framework allows us to define the device behaviour and operations in clear steps and use the interface templates to implement the behaviour in a way that is understandable for other components of the software.

On the other hand, the short-duration of balloon-flights, compared to longer-duration space missions, requires an efficient scheduling for each mission. In addition, given that the aim of the ESBO is to accommodate several observation blocks (which may originate from different observers) in flight, there would be a need to understand and coordinate the requirements (pointing, observation fields, observation durations, and similar) for different each one, and to keep the operations and schedule as flexible and as autonomous as possible, using the lessons learned from previous ground-based and space-based missions.

In this case, the implementation of PUS time-based scheduling is currently used, including the features to have subschedules and modify these, and by adding a pre-loader interface to set up an on-board mission schedule at software start up. This implementation will be used in future satellite missions at the Institute of Space Systems / University of Stuttgart as well, and it will be developed further for future ESBO missions.

To take this one step further is to implement an efficient autonomous on-board event-driven rescheduling procedures, to react to changes in the environment (e.g. unplanned changes in altitude, which might make an observation impossible to carry on), or on-board failures (failure to perform one step of an observation).

In addition, our future goal is to complete the autonomous scheduling chain by creating a possibility for instrument teams to communicate their observation plans, or change in plans, to the mission control system using observation scripts that are parsed within the mission control system. This flexible multi-instrument scheduling can be a potential spin in for space missions as well.

## 6. Critical technologies for the way forward

In order to take further steps from the STUDIO prototype to a regularly flying balloon observatory, particularly also including the FIR capabilities mentioned in chapter 3, several technologies are critical. The most important ones shall be shortly summarized in the following.

### 6.1 Lightweight, stabilized balloon gondolas

Given that there is a practical limit to the maximum payload liftable to a certain altitude, particularly for larger telescopes and heavier payloads, lightweight balloon gondolas are vital. For astronomical applications, they furthermore need to be stabilized at least to the degree where an internal image stabilization system can pick it up. This is typically at a few tens of arcseconds or less. STUDIO is taking some steps of this development already, however, to achieve further lightweighting, it will be inevitable to look into CFRP gondola structures, combined with more controlled landing techniques than currently applied.

### 6.2 Reliable highly accurate image stabilization

The image stability required for modern-day astronomical observations can hardly be achieved by a balloon gondola pointing system alone. Two-stage system, employing an additional image stabilisation system in the optical system, seem much more practical. STUDIO is taking an important step by implementing such a system for its application in the UV and visible. For an FIR application this system will need to be adjusted and made daytime compatible.

### 6.3 Large CFRP mirrors

The feasibility of the 5 m-aperture FIR concept relies upon the possibility to obtain a light-weight mirror of this size, for which CFRP structures seem the most promising. A critical development step is therefore the demonstration of a 5 m CFRP mirror, capable of maintaining thermomechanical stability under the flight conditions, and with an areal density of $20 \, kg/m^2$ or less.

### 6.4 Higher payload mass capacity balloons

The free lift of currently existing ZPBs is actually adequate for lifting the mass of a large FIR observatory to 20 to 30 km altitude. Past balloons have also carried masses of up to 10 t to lower altitudes. The design of current balloons will need to be adjusted to structurally allow the higher suspended mass, however.

### 6.5 Safe landing technology tests

One of the main concerns for reflight of ballooning hardware is the current landing technique, which relies upon unsteered parachute landings. For a regularly flying observatory, more controlled, steered landings would bring a great improvement of efficiency. A suitable landing technique is presented by autonomous, steered landing systems based on steerable parafoils which are commercially available for military and humanitarian applications. Their applications in a two-stage system used on a stratospheric balloon still needs to be demonstrated, however.

## Acknowledgements
ESBO *DS* has received funding from the European Union's Horizon 2020 research and innovation programme under grant agreement No 777516.